\documentclass[preprint,aps]{revtex4}

\begin{document}
\date{\today}

\title
{Bose-Einstein condensation of magnons}

\author
{M. Crisan, D. Bodea, I. Tifrea and I. Grosu}

\address
{Department of Theoretical Physics,University of Cluj, 3400 Cluj,
Romania}

\begin{abstract}
We use the Renormalization Group method to study the Bose-Einstein
condensation of the interacting dilute magnons which appears in
three dimensional spin systems in magnetic field. The obtained
temperature dependence of the critical field $H_c(T)-H_c(0) \sim
T^{2}$ is different from the recent self-consistent
Hartree-Fock-Popov calculations (cond-matt 0405422) in which a
$T^{3/2}$ dependence was reported . The origin of this difference
is discussed in the framework of quantum critical phenomena.
\end{abstract}

\maketitle

\newpage

\section{Introduction}

Recently, the quantum spin system in magnetic field became of
great interest because the experimental data on the ladder systems
showed the possibility of a quantum phase transition (QPT) driven
by the magnetic field\cite{1}. The QPT in the spin systems have
been treated using the quantum rotor model\cite{2}. However, the
transition driven by a magnetic field is difficult to be described
in such a formalism.

The occurrence of the magnetic order in the spin-gap magnetic
compounds has been interpreted as a Bose-Einstein condensation
(BEC) of magnons. At the present time it is known that the BEC has
been discovered in ultra-cooled dilute atomic gases, but these
experiments present various limitations. The analogy between a
quantum spin system which presents long-range order and an
interacting Bose gas which presents BEC is well known for a long
time\cite{3}. In the case of quantum spin system is possible to
tune the density of magnons by magnetic field \cite{11,12} or
pressure \cite{20} to observe BEC in these systems.

In this paper we formulate the theory of magnon condensation using
the Renormalization Group method (RNG) applied for bosonic
systems\cite{4,5,6,7,8,9} and recently reconsidered for the spin
systems\cite{10}.

The outline of this paper is as follows. In Sect. II we present
the basic experimental evidence obtained on the material class
$XCuCl_3 \; (X=Tl,K,NH_4)$ and the main results obtained with the
mean field theory. The model based on these data will be presented
in Sect. III. The Sect. IV is devoted to the RNG formulation using
results from Ref. \cite{10} in order to calculate the low
temperature quantum critical properties of the system. Sec. V will
present the main results of the paper, the relevant thermodynamic
quantities next to the quantum critical point. Finally, in the
last section we will discuss the results and compare with other
theoretical approaches and experimental evidence.

\section{Experimental evidences}

In this section we present the basic experimental data obtained on
the $XCuCl_3$ compounds which gave us the possibility to elaborate
a model for the critical behavior of this QPT as a BEC driven by
the magnetic field.

The compound which offered the possibility of performing many
measurements is $TlCuCl_3$ which is composed of a chemical double
chain $Cu_2Cl_6$. The magnetic susceptibility, measured as
function of temperature T for three different directions of the
magnetic field, exhibit a broad maxima at $T=38K$, decreasing to
zero with the temperature decreasing \cite{11}. This result
indicates that the compound has an excitation gap
$\frac{\Delta}{k_B}=7.5K$ above the singlet ground state. This gap
may by attributed to the antiferromagnetic dimer coupling in the
double chain \cite{12}. The magnon dispersion has been
investigated theoretically \cite{13} and experimentally by
inelastic neutron scattering \cite{14}. The experiments showed
that the magnon modes are split into three by magnetic field with
the splittings proportional to the field , and the lower modes
becomes soft at the critical field $H_c$. In a magnetic field $H$
the chemical potential is $\Delta-g\mu_BH$ (where $g=2$) and
$H_c=\frac{\Delta}{\mu_Bg}$.

An important result has been obtained by Sherman {\it et al.}
\cite{15} by sound attenuation in $TlCuCl_3$ in magnetic field at
low temperatures. The occurence of the sharp peak in the sound
attenuation near the BEC critical temperature and the Drude form
of the sound dumping suggested a constant weak magnon- magnon
coupling.

The sound attenuation suggested a magnons dispersion like
$\omega^2=\Delta^2+\frac{J^2k^2}{4}$ (here $J$ is the exchange
interaction) for small wave vector "$k$", approximation valid if
$T_c$ and $\Delta$  are of the order of $k_BT_c=7K$ and
$k_B\Delta=7.5$ as for the case of $TlCuCl_3$.

The existence of a induced magnetic field-spin transition has been
proposed by Giamarchi and Tsvelik \cite{16} considering as a
possible mechanism the BEC of the soft mode. The Hartree-Fock
approximation using the Hamiltonian from the theory of BEC in
dilute bosonic gases with effective chemical potential
$\mu=\mu_Bg(H-H_c)$ has been applied \cite{17} to calculate the
temperature dependence of magnetization. Near the critical
temperature $T_c$ this approximation breaks down and the
magnetization is expected to behave like $M\sim(T-T_c)^\beta$
where $\beta=\frac{3}{2}$. The temperature dependence of the
critical field has the form $[H_c(T)-H_c(0)]\sim T^\Phi$ where
$\Phi=\frac{3}{2}$ but the best fit is obtained for $\Phi=2.2$.

The system has some important characteristics. First, we menton
the small three-dimensional (3D) character of the system which
leads to a small gap. Another important feature is the decreasing
the number of thermal excitations with the increasing of the
magnetic field, and as a consequence the driving the condensate to
$T=0$ behavior.

The form of the magnon dispersion (called "relativistic") changes
the dynamics of the system and we expect $\Phi=2$. This conjecture
is on of the most important result of Ref. \cite{15} and we will
use it for the model of the magnon condensation.

\section{Microscopic model}

We introduce the description of the magnon condensation in
magnetic field using the action:
\begin {equation}
  S_{eff}\;=\;S_{eff}^{(2)}\;+\;S_{eff}^{(4)}
\label{1}
\end {equation}
where:
\begin{equation}
S_{eff}^{(2)}\;=\frac{1}{2}\;\sum_k\;\chi^{-1}(k)|\phi(k)|^2
\label{2}
\end{equation}
\begin{equation}
S_{eff}^{(4)}\;=\frac{u_{0}}{16}\sum_{k_1}\ldots\
\sum_{k_4}\;\phi(k_1)\ldots\phi(k_4)\delta(k_1+\dots+k_4)
\label{3}
\end{equation}
Here we introduced the notations  $k=({\bf k },{\omega_n})$,
$\omega_n$ being the bosonic Matsubara frequency and
\begin{displaymath}
\sum_k = T\sum_n \int \frac{d^d {\bf k}}{(2\pi)^d}
\end{displaymath}

In Eq.(\ref{2}) $\chi(k)$ is the magnon propagator and $u_{0}$ the
bare coupling constant. Using \cite{15} we write the magnon
propagator as
\begin {equation}
\chi({\bf k},{\omega_n})=\;\frac{1}{\omega_n^2+\bf
k^2+r_0}
\label{4}
\end{equation}
where $r_0=\Delta-\mu_BgH$ and H is the external magnetic field.

 \section{Renormalization group equations}

The model has the dynamical critical exponent $z=1$ and for this
model $d=3$. The Renormalization group equations in one-loop
approximation at finite temperature $T$ for $d=3$ have the form
\cite{10}:
\begin{equation}
\frac{dT(l)}{dl}\;=\;T(l)
\label{5}
\end{equation}
\begin{equation}
\frac{du(l)}{dl}\;=\;-\frac{(n+8)K_{3}}{8} u^{2}(l)\;
F_{1}[r(l),T(l)]
\label{6}
\end{equation}
\begin{equation}
\frac{dr(l)}{dl}\;=2r(l)\;+\frac{(n+2)K_{3}}{8} u(l)\; F_{2}[r(l),
T(l)]
\label{7}
\end{equation}
where $n$ is the number of components of the fluctuation field
$\Phi(k)$,  $K_{3}$=$\frac{1}{2\pi^2}$. We denote by
$u(l=0)=u_{0}$,and $r(l=0)=r_{0}$ and the functions $F_{1,2}$ are
characteristic functions for the model \cite{10}. We approximate
$F_{2}[r(l), T(l)] \simeq\frac{1}{4}$ in the low temperatures
limit and the solution of Eq. (\ref{5}) has the form:
\begin{equation}
 u(l)\;=\;\frac{1}{C_{0}(l+l_{0})}
\label{8}
\end{equation}
where  $C_{0}= \frac{(n+8)K_{3 }}{16}$  and
$K_{3}=\frac{1}{2\pi^2}$. The physics near the QCP is described by
the scaling field $t_{r}(l)$ defined as:
\begin{equation}
t_{r}(l)=r(l)+\;\frac{n+2}{16}K_{d}u(l)
\label{9}
\end{equation}
The expression of $t_{r}(l)$ has been calculated in Ref. \cite{10}
for d=3 as:
\begin{equation}
 t_{r}(l)=  e^{\Lambda_{r}(l)} \left\{t_{r}(0)+\frac{n+2}{4}K_{d}\int_{0}^{l}dx \;
 \frac{e^{-2x}u(x)}{e^{1/T(x)}-1}\right\}
\label{10}
\end{equation}
where $\Lambda_{r}(l)$ has the expression:
\begin{equation}
\Lambda_{r}(l)=2l-\frac{n+2}{n+8}\ln(\frac{l}{l_{0}}+1)
\label{11}
\end{equation}
Using these results we will calculate the thermodynamic quantities
in near the critical point. The basic idea of the method is to
stop the renormalization procedure close enough to this point that
the system can see the influence of the quantum effects. This
matching condition is in fact equivalent to the stoping of the
renormalization at the scale $l=l^{*}$,where $l^{*}\gg 1$. This
value is obtained from the condition \cite{10}:
\begin{equation}
t_{r}(l^{*})=1
\label{12}
\end{equation}

In the approximation $l^{*}>1$ and $T(l^{*})>1$ we obtain from
Eq.(\ref{11}) the relation:
\begin {equation}
\exp({l^{*}})=[t_{r_{0}}(T)]^{\frac{-1}{\lambda_{r}}}
\label{13}
\end{equation}
where $\lambda_r$ is the eigenvalue of the relevant parameter $r$
and is given by the expression:
\begin{equation}
\lambda_{r}=2,    \epsilon=0     \;(d=3)
\label{14}
\end{equation}
and $t_{r_{0}}(T)$ is given by the relation \cite{10}:
\begin{equation}
t_{r_{0}}(T)=r_{0}-r_{0c}+\frac{n+2}{128}u_{0}T^{2}
\label{15}
\end{equation}
Following the method from \cite{7,8} we calculate $l^*$ as:
\begin{equation}
l{^*}=\frac{1}{2} \ln\frac{1}{T}
\label{16}
\end{equation}

\section{Thermodynamic quantities}

Using these results we will calculate the relevant thermodynamic
quantities near the quantum critical point, but in the disordered
state. First, we define the critical  line by $t(T)=0$ and we get:
\begin{equation}
H_c(T)-H_c(0)=C_{0}T{^2}
\label{17}
\end{equation}
where $C_{0}\propto{u_{0}}$ and $H_{c}=\frac{\Delta}{\mu_{0}g}$.
The temperature dependence of the number of magnons is defined as:
\begin{equation}
n(T)=exp  {(-3 l{^*})}\int\frac{d{^{3}\bf
k}}{(2\pi)^{3}}f_{B}[T(l{^*})]
\label{18}
\end{equation}
This equation gives for $n(T)$ a temperature dependence of the
form:
\begin{equation}
n(T)\propto{T{^\frac{3}{2}}}
\label{19}
\end{equation}
This result is in agreement with the behavior of magnetization at
very low temperatures.

From Eq. (\ref{15}) we calculate the critical line in the
$(r_0,T)$ plane using the condition $t_{r_0}(T)=0$. This gives:
\begin{equation}
r_{0c}(T)=r_{0c}-\frac{(n+2)}{128}u_0 T^2
\label{20}
\end{equation}
and for $r_0 \le r_{0c}$ we get the general equation
\begin{equation}
T_c(r_0)=\left[ \frac{128}{(n+2)u_0} \right]^{1/2}
(r_{0c}-r_0)^{1/2}
\label{21}
\end{equation}
Using now the definition (see Ref. \cite{10}) of $r_{0c}$
\begin{equation}
r_{0c}=- \frac{(n+2)}{32 \pi^2}u_0
\label{22}
\end{equation}
we obtain
\begin{equation}
T_c(H) \sim |H-\tilde{H}_c|^{1/\alpha}
\label{23}
\end{equation}
with $\alpha=2$ and
\begin{displaymath}
\tilde{H}_c = H_c + \frac{(n+2)}{32 \pi^2} u_0
\end{displaymath}

The specific heat can be also calculated for the important case
$r_0 \ne r_{0c}$ but $T \rightarrow T_c^+(r_0)$ using the singular
part of the free energy reported in \cite{10} as:
\begin{equation}
F_s(T) \sim [T - T_c(H)]^2 \; |\ln(T-T_c(H))|^{\left[ 2
\frac{n+2}{n+8} \right]}
\label{24}
\end{equation}
and it give us a logarithmic behavior:
\begin{equation}
\frac{C_v(T)}{T} \sim |\ln(T-T_c(H))|^{\left[ -2\frac{n+2}{n+8}
\right]}
\label{25}
\end{equation}
We will discuss these results in connection with the existent
theoretical approaches and the experimental data for $Tl Cu Cl_3$.

\section{Discussions}

The occurence of BEC induced by the magnetic field in a spin-gap
system has been predicted by Giamarki and Tsvelik \cite{16} in
order to explain the three-dimensional ordering in coupled
ladders. Nikumi {\it et al.} \cite{17} applied Popov theory (see
for example \cite{18}) for the BEC obtaining a temperature
dependence for magnetization in agreement with the experimental
data, of the form $T^{3/2}$.

However, as it was mentioned in Ref. \cite{17} the critical
exponent $\Phi$ of the critical field $[H_c(T)-H_c(0)] \sim
T^{\Phi}$ was obtained as $\Phi=3/2$, but the experimental data
shows $\Phi=2.2$.

Recently Misguich and Oshikawa \cite{19} reconsidered the
self-consistent Hartree-Fock-Popov (HPF) method using a realistic
dispersion for magnons \cite{13} and they reobtained $\Phi=3/2$.
They also calculated the specific heat and obtained a
$\lambda$-shape.

We would like to mention several basic points related to this
problem:
\begin{itemize}
\item in the HPF the effect of fluctuations have been neglected
and it appears normal that thermodynamic quantities on the
critical line have wrong critical exponents.
\item the changing of the dispersion law for the magnon can
improve the HFP results, but does not affect the basic approach.
\item both papers \cite{17} and \cite{18} considered the QPT as
driven by the temperature and in fact it is induced by the
magnetic field.
\end{itemize}
These observations justify the application of the RNG method in
the version proposed by Caramico {\it et al.} \cite{10} to study
the magnon condensation. Our main result can be summarized as:
\begin{itemize}
\item we obtained the critical exponent $\Phi=2$, which is in a
better agreement with experimental data ($\Phi=2.2$).
\item the magnetization calculated in our approach is also
$T^{3/2}$ dependent.
\item the critical temperature $T_c(H)$ has a dependence given by
Eq. (\ref{23}) with $\alpha=2$ in agreement with \cite{12}. The
numerical results have been analyzed in \cite{20}.
\item the specific heat calculated by RNG method has a logarithmic
correction as was expected. The $T$ dependence of $C_v$ is not
relevant if we take the approach of a magnetic field driven QPT,
but we have it to the dependence of H(T). Such an experiment has
been done \cite{12} and we can notice a small anomaly in this
dependence. But a final decision about the $H(T)$ dependence of
our Eq.(25) cannot be taken.
\end{itemize}
The agreement between the experimental data and the theory
presented in \cite{19}appears in our opinion due to the fact that
the authors considered, on a very narrow interval a linear
dependence between the critical field $H_{c}(T)$ and the critical
density$n_{c}$. This approximation can be valid for $n_{c}<0.002$
which describes a part of the real temperature dependence of
$H_{c}(T)$.

\section{Acknowledgments}
One of the authors (M. C.) thanks Luigi De Cesare for numerous
correspondences helping him to understand the RG methods developed
by the Salerno group(A. Caramico D'Auria, L. De Cesare and I.
Rabuffo) for this problem.

\end{document}